\documentclass[conference]{IEEEtran}

\makeatletter

\usepackage{blindtext}
\usepackage{eso-pic}
\usepackage{url}
\usepackage{hyperref}
\IEEEoverridecommandlockouts
\usepackage{cite}
\usepackage{amsmath,amssymb,amsfonts}
\usepackage{algorithmic}
\usepackage{graphicx}
\usepackage{textcomp}
\usepackage{xcolor}
\def\BibTeX{{\rm B\kern-.05em{\sc i\kern-.025em b}\kern-.08em
    T\kern-.1667em\lower.7ex\hbox{E}\kern-.125emX}}
    
\usepackage{eso-pic}

\begin{document}
\title{CLASP: Cost-Optimized LLM-based Agentic System for Phishing Detection}

\author{\IEEEauthorblockN{Fouad Trad}
\IEEEauthorblockA{\textit{Electrical and Computer Engineering} \\
\textit{American University of Beirut}\\
Beirut, Lebanon \\
fat10@mail.aub.edu}
\and
\IEEEauthorblockN{Ali Chehab}
\IEEEauthorblockA{\textit{Electrical and Computer Engineering} \\
\textit{American University of Beirut}\\
Beirut, Lebanon \\
chehab@aub.edu.lb}

}

\maketitle

\begin{abstract}
Phishing websites remain a significant cybersecurity threat, necessitating accurate and cost-effective detection mechanisms. In this paper, we present CLASP, a novel system that effectively identifies phishing websites by leveraging multiple intelligent agents, built using large language models (LLMs), to analyze different aspects of a web resource. The system processes URLs or QR codes, employing specialized LLM-based agents that evaluate the URL structure, webpage screenshot, and HTML content to predict potential phishing threats. To optimize performance while minimizing operational costs, we experimented with multiple combination strategies for agent-based analysis, ultimately designing a strategic combination that ensures the per-website evaluation expense remains minimal without compromising detection accuracy. We tested various LLMs, including Gemini 1.5 Flash and GPT-4o mini, to build these agents and found that Gemini 1.5 Flash achieved the best performance with an F1 score of 83.01\% on a newly curated dataset. Also, the system maintained an average processing time of 2.78 seconds per website and an API cost of around \$3.18 per 1,000 websites. Moreover, CLASP surpasses leading previous solutions, achieving over 40\% higher recall and a 20\% improvement in F1 score for phishing detection on the collected dataset. To support further research, we have made our dataset publicly available, supporting the development of more advanced phishing detection systems.
\end{abstract}


\section{Introduction}

Phishing attacks remain one of the most prevalent and damaging cybersecurity threats, targeting individuals, organizations, and critical infrastructure alike. These attacks often take the form of fraudulent websites that impersonate legitimate services to deceive users into revealing sensitive information such as passwords, financial details, and personal data. As phishing tactics evolve, the methods employed to detect and mitigate such threats must also advance. Traditional phishing detection techniques often rely on a combination of heuristic analysis and URL inspection, but the growing sophistication of phishing websites demands more advanced, resource-efficient solutions that can adapt to emerging attack strategies \cite{da2020heuristic}. Machine learning-based solutions have proven effective in addressing some of these challenges by identifying complex patterns and adapting to new threats. However, they require continuous training and maintenance to remain effective against evolving phishing tactics \cite{tang2021survey}. Prompt-engineered large language models (LLMs) offer an alternative by leveraging their extensive pretraining and adaptability through carefully designed prompts, eliminating the need for model retraining \cite{trad2024ensemble}. While this approach enhances phishing detection flexibility and reduces maintenance overhead, it introduces the challenge of API costs, as LLM-based systems rely on external providers for processing \cite{trad2024large}. Additionally, since these models are not fine-tuned specifically for phishing detection, they may not perform as well as dedicated machine learning models \cite{trad2024evaluating,trad2024prompt}. The challenge, therefore, lies in designing effective prompting strategies that maximize their performance and reliability in detecting phishing websites \cite{trad2025manual}.

In this paper, we propose CLASP, a novel system that utilizes multiple intelligent agents, built on LLMs, to detect phishing websites. Our system processes various aspects of a website, including the URL, webpage screenshot, and HTML content, and employs specialized agents to analyze each of these components individually. The goal is to provide accurate and cost-effective phishing detection by leveraging the unique strengths of LLMs to evaluate different facets of a web resource. Since LLMs are used through prompting rather than training or fine-tuning, and their providers continuously update them, our system eliminates the need for model retraining while still adapting to evolving phishing tactics. By combining the results of these agents, we aim to achieve improved performance without significantly increasing API costs.

To establish a baseline, we first tested individual agents that analyzed only one component at a time: the URL, the webpage screenshot, or the HTML content. Each of these methods, while useful on its own, exhibited limited performance in terms of detection accuracy. This was expected, as phishing websites often exhibit a combination of cues across different components that, when analyzed in isolation, may not provide sufficient information to identify them accurately.

Next, we explored different methods for combining the results of the individual agents. We initially experimented with majority voting, where the final prediction was based on the most common result across all agents. However, this approach did not yield the desired performance. We then tested a sequential processing strategy, where agents analyze the website progressively, starting with the URL, followed by the webpage screenshot, and finally the HTML content. If any agent flags the website as phishing at any stage, the system immediately classifies it as phishing, eliminating unnecessary calls to subsequent agents. Our findings revealed that the progressive combination strategy not only provided the best detection performance but also maintained a low operational cost. On average, this method processed a website in just 2.78 seconds, with a cost of \$3.18 per 1,000 websites. This combination strategy represents an effective trade-off between accuracy and resource efficiency, making it a promising solution for large-scale phishing detection applications.

Evaluated on a newly curated dataset, CLASP outperforms existing commercial phishing detection solutions by over 40\% in recall and 20\% in F1 score, demonstrating the potential of leveraging multiple LLM-based agents for enhanced phishing detection. To foster further research and development in this area, we have made our dataset publicly available, enabling other researchers to build upon and improve our approach.

In summary, the main contributions of this work are:

\begin{enumerate}

    \item Development of CLASP, a multi-agent phishing detection system that analyzes multiple components of a phishing website, including the URL, webpage screenshot, and HTML content, to improve detection performance.
    \item Evaluation of different agent-based techniques for combining agent outputs.
    \item Balancing detection accuracy with resource efficiency, achieving an average processing time of 2.78 seconds per website and a cost of \$3.18 per 1,000 websites, making our approach highly cost-effective for large-scale deployment.
    \item Outperforming leading commercial phishing detection products by more than 20\%, showcasing the potential of leveraging multi-agent LLM-based approaches in real-world applications.
    \item Curating a new phishing dataset and making it publicly available to support further research and development in this area.
\end{enumerate}

The remainder of the paper is organized as follows: Section 2 provides the background and preliminaries essential to our study. Section 3 reviews related work. Section 4 outlines the methodology, while Section 5 details the experimental setup and presents the results, discussing the findings from both performance and cost perspectives. Finally, Section 6 concludes the paper and suggests directions for future research.

\section{Background and Preliminaries}
\subsection{Large Language Models (LLMs)}
LLMs are advanced artificial intelligence systems designed to interpret and generate text that closely resembles human language. These models are primarily built on transformer architectures, which enable them to process extensive textual data and uncover intricate language patterns \cite{zhao2023survey}. Well-known examples of LLMs include OpenAI's GPT models \cite{roumeliotis2023chatgpt} and Google's BERT \cite{devlin2018bert}. LLMs have demonstrated exceptional performance in various natural language processing (NLP) tasks, such as text generation, language translation, and answering questions, due to their ability to utilize vast datasets and sophisticated training methodologies \cite{chang2024survey}.

\subsection{Agentic AI Systems}
Agentic AI systems involve the collaboration of multiple autonomous agents working together towards a common objective. In the world of LLMs, these multi-agent systems can enhance the overall performance of the models by allowing specialized agents to focus on different aspects of input data \cite{xie2024large}. For instance, one agent may process linguistic features, another might analyze visual elements, and yet another could manage contextual understanding. By distributing tasks among multiple agents and enabling them to communicate and coordinate, these systems can deliver more accurate and efficient results than relying on a single, monolithic model \cite{sreedhar2024simulating}. However, creating such systems presents challenges, including ensuring seamless communication, coordination, and trust among agents. Effective collaboration is essential for ensuring that the agents align their efforts toward the system's overarching objectives and make decisions that benefit the overall application.

\section{Related Work}
Phishing detection has evolved significantly over the years, with early methods relying heavily on traditional heuristic and rule-based approaches \cite{qabajeh2018recent, da2020heuristic}. These systems often focused on identifying known malicious URLs or applying keyword-based filters to detect phishing websites \cite{ding2019keyword, tan2016phishwho}. While these foundational methods provided a starting point, they were often prone to high false-positive rates and struggled against more advanced phishing attempts that closely mimic legitimate websites or communications \cite{ding2019keyword}.

As machine learning began to play a larger role in phishing detection, more advanced models such as decision trees \cite{machado2017phishing}, support vector machines \cite{altaher2017phishing}, and ensemble methods \cite{ubing2019phishing} emerged. These models leveraged large datasets to identify patterns in phishing URLs, which improved their detection accuracy. Deep learning techniques, including convolutional neural networks (CNNs) and recurrent neural networks (RNNs), were also applied to analyze more complex data, such as images and natural language \cite{alshingiti2023deep, xiao2020cnn}. These deep learning models generally outperformed traditional methods, achieving higher accuracy in phishing detection. However, they required substantial computational resources and continuous retraining, making them less feasible for real-time or large-scale deployment.

In recent years, the use of LLMs in cybersecurity applications has gained significant attention \cite{zhang2024llms, motlagh2024large}. Prompt-engineered LLMs, in particular, have demonstrated considerable potential for phishing detection due to their ability to adapt dynamically using specific input prompts, without the need for ongoing retraining \cite{trad2024prompt, trad2024ensemble}. These models have been applied to analyze textual data from phishing URLs \cite{ferrag2024generative}, offering improvements over traditional machine learning approaches.

While the application of LLMs to phishing detection is promising, previous studies have primarily focused on analyzing a single modality, such as the URL alone, or the HTML content of websites. However, the integration of LLMs with multiple agents to process different aspects of a website for phishing detection remains largely unexplored. Furthermore, while research on LLMs in phishing detection has gained traction, there is limited work on evaluating the cost of these models, particularly regarding the computational and financial costs associated with API-based systems \cite{wang2024reasoning,trad2024large}.

This work introduces a novel multi-agent LLM-based system that analyzes multiple aspects of phishing websites, including URLs, HTML content, and screenshots, to improve detection accuracy. Unlike traditional approaches that rely on a single type of data, our system combines the strengths of multiple agents, each specialized in a different website component, to make more informed predictions. Moreover, the system is designed to optimize both detection performance and cost-efficiency, addressing the economic challenges associated with API usage by minimizing the number of queries required for each website.

\section{Methodology}
The proposed system leverages three LLM-based agents that analyze distinct data representations (URL, Screenshot, and HTML) of a given website to predict phishing attempts. The Screenshot Agent inspects the visual presentation to identify deceptive branding or misleading elements commonly seen in phishing websites. The URL Agent analyzes the URL’s structure and characteristics to assess its potential as a phishing link. The HTML Agent evaluates the webpage’s content to detect suspicious patterns, malicious scripts, or phishing indicators. The process begins with the system accepting either a URL or a QR code as input. If a QR code is provided, it is first decoded to extract the corresponding URL.

\begin{figure*}[ht!]
    \centering
    \begin{minipage}{0.48\textwidth}
        \centering
        \includegraphics[width=\textwidth]{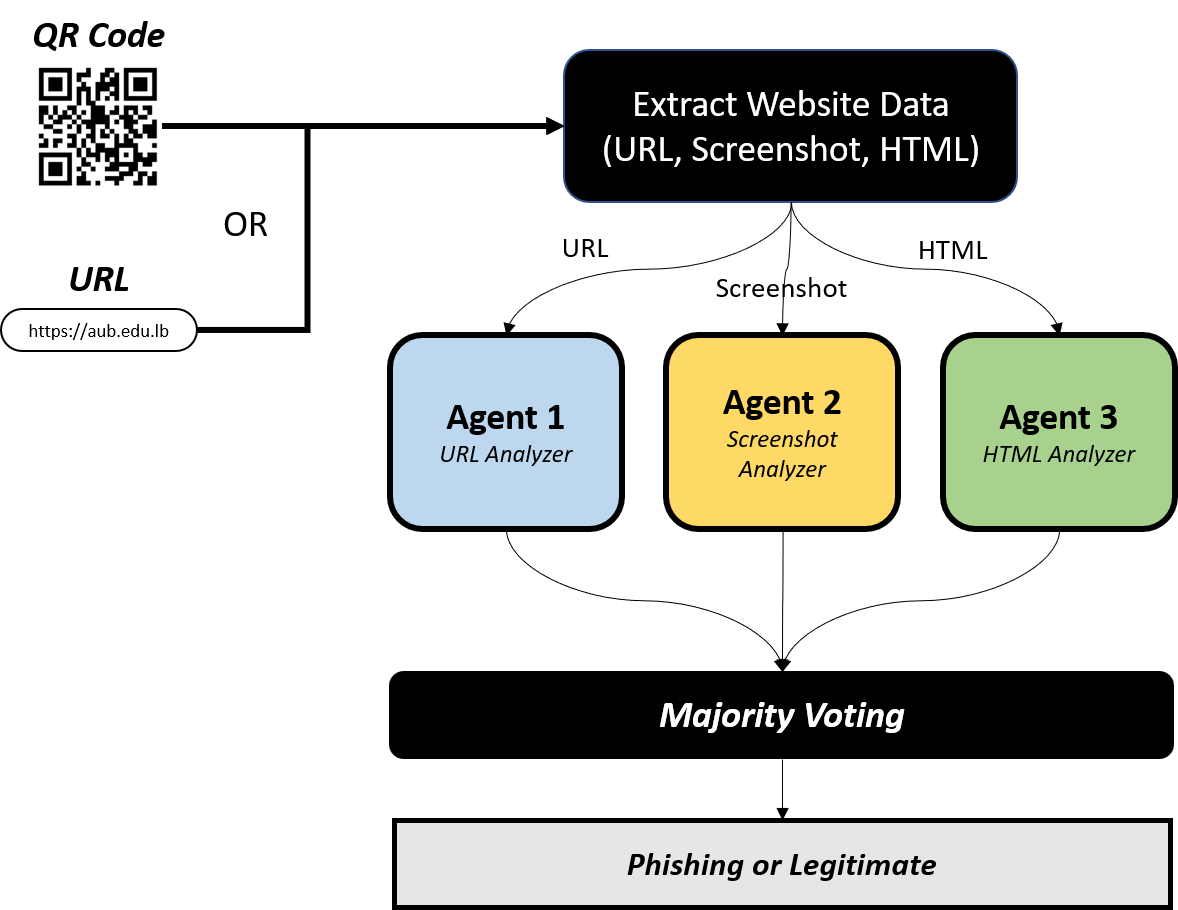} 
    \end{minipage}
    \hspace{0.05cm}
    \begin{minipage}{0.48\textwidth}
        \centering
        \includegraphics[width=\textwidth]{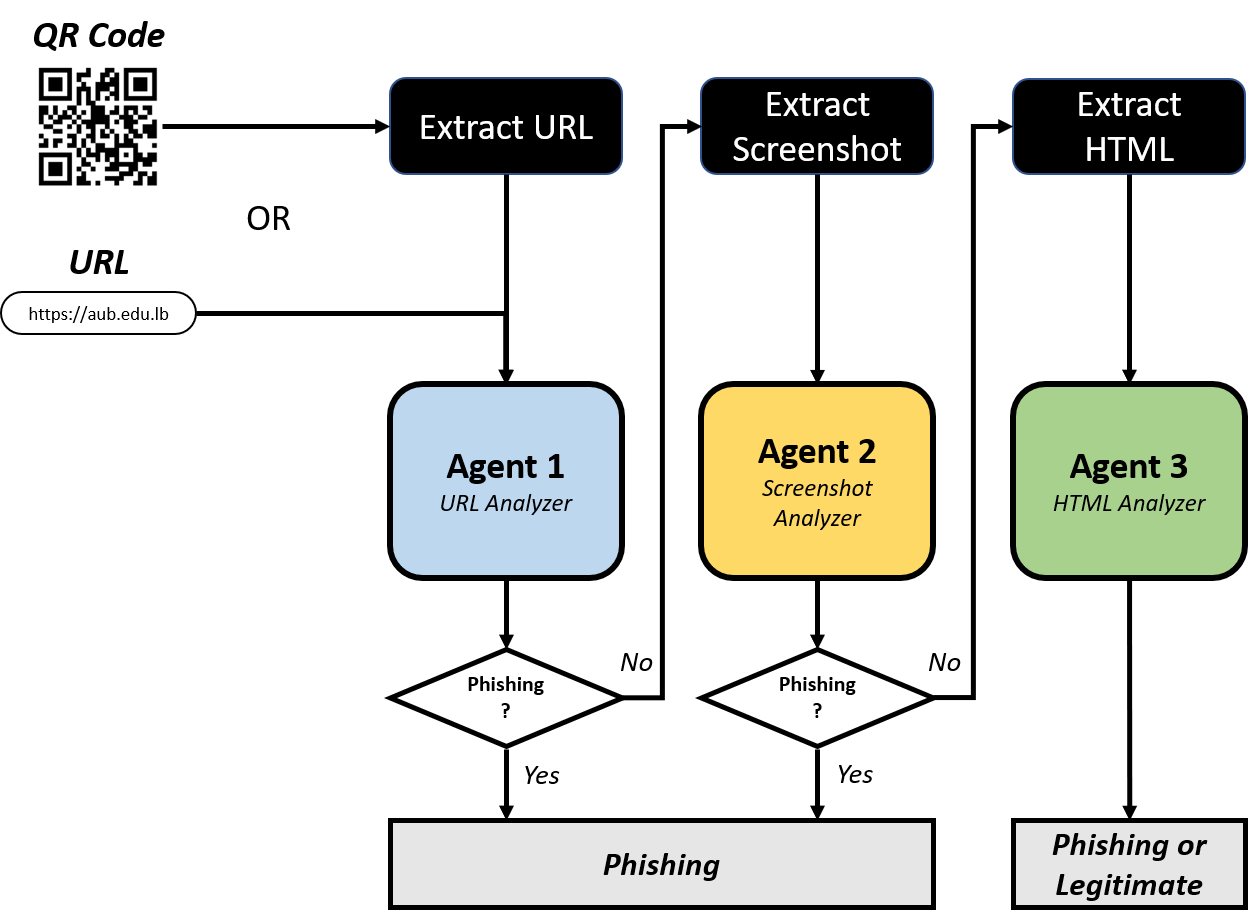}
    \end{minipage}
    \caption{Comparison of (a) Majority Voting and (b) Progressive Analysis.}
    \label{fig:comparison}
\end{figure*}

Initially, we tested each agent on its own to evaluate their baseline performances. Then, we explored multiple strategies to combine the agents' outputs. We started with a Majority Voting Strategy, where a phishing label was assigned if at least two out of the three agents predicted phishing. The process is summarized in Figure \ref{fig:comparison} (a). This method aims to improve detection robustness by ensuring that phishing identification relies on consensus among the agents. This strategy incurs higher operational costs due to the need to always extract and analyze all three data types: URL, screenshot, and HTML content.

To improve cost efficiency, we adopted a Progressive Analysis Strategy. This method starts by analyzing the URL alone through the URL Agent. If the URL agent does not detect phishing, the system proceeds to extract a full-page screenshot for analysis by the Screenshot Agent. If this agent also fails to identify phishing characteristics, only then the full HTML content is extracted and processed by the HTML Agent. The process is summarized in Figure \ref{fig:comparison} (b). This progressive approach minimizes resource consumption by limiting data extraction steps and agent analyses to what is necessary.

To assess these methods, we measured detection performance using metrics such as accuracy, F1 score, precision, and recall. Additionally, we evaluated cost efficiency by calculating the average API expenses and monitored the time taken to analyze each website to ensure practical usability.

\section{Experiments}
\subsection{Experimental Setup}
\subsubsection{Data}
To evaluate our system, we utilized a subset of the PhiUSIIL dataset, which comprises recently collected websites labeled as legitimate or phishing \cite{prasad2024phiusiil}. For each website, we extracted its screenshot and HTML content; these extracted data elements will be made publicly available to support future research. The dataset can be accessed \href{https://www.dropbox.com/scl/fi/2z5xd4tdebd7c3fl2nzhv/PhishingDataset.zip?rlkey=vt6zzg3wujz0076uhhwx3wnvq&st=1i8v49qf&dl=0} {here}. Due to the nature of phishing websites, some URLs were inaccessible at the time of data collection (either because the phishing site had been taken down or due to server unavailability) resulting in a final sample of 2,126 websites (1,110 legitimate and 1,016 phishing).

\subsubsection{Models}
For models, we began by building our agents using Gemini 1.5 Flash. After identifying the best-performing setup, we evaluated it on GPT-4o mini to gain insights into how different models perform in this context. These models were selected for their accessibility via API and their relatively lower costs compared to other models, making them particularly viable for industry applications that aim to mitigate phishing threats while maintaining manageable operational costs.

\subsubsection{Prompts}
We employed structured output prompts designed to instruct the models to act as "cybersecurity professionals" investigating different website components. Each prompt directed the model to classify the content as either "Phishing" or "Legitimate" while providing a succinct reasoning for the decision. This reasoning is essential for helping users understand why a given website is flagged as phishing or deemed safe. For example, the prompt used for the URL agent was:

\textit{"Act as a professional cybersecurity analyst reviewing URLs. Classify the following URL as Phishing or Legitimate. Your reasoning should be succinct"}

Similar prompts were adapted for the screenshot and HTML agents, ensuring each agent's task was clearly defined to maximize performance. 

\subsection{Assessing Baseline Agents}
In this initial experiment, we evaluated the performance of individual agents that analyze distinct website components: URL, screenshot, and HTML content.

The results, shown in Table \ref{tab:performance_experiments}, demonstrate that the URL Agent achieved the strongest overall performance among the individual agents, with an accuracy of 0.799 and an F1 score of 0.773. This agent's relatively balanced precision (0.840) and recall (0.717) enabled it to detect a reasonable portion of phishing websites without excessive false positives.

The Screenshot Agent and HTML Agent achieved even higher precision values (0.884 and 0.967, respectively), but their recall rates were notably lower (0.457 and 0.483). This reveals that while these agents excel at confirming phishing attempts when identified, they are less effective at detecting harder-to-identify phishing websites, reducing their overall reliability as standalone solutions.

These results highlight a common challenge in phishing detection: while achieving high precision is important for avoiding false positives, poor recall significantly limits the system's ability to capture a wider range of phishing threats.

Table \ref{tab:resources} presents the resource consumption of different agents, with times referring only to the analysis phase and not including the time required for capturing screenshots or saving HTML content. In terms of resource efficiency, the URL Agent proved to be the fastest and most cost-effective, processing a URL in 0.68 seconds at a cost of \$0.018 per 1,000 websites. The HTML Agent, while achieving higher precision, required significantly more resources, averaging 2.28 seconds per website at \$4.697 per 1,000 websites. The Screenshot Agent had an intermediate processing time of 1.18 seconds. These findings highlight the need for strategies that optimize both detection performance and resource efficiency.

\subsection{Majority Voting}
In this experiment, we explored a Majority Voting strategy, where a website is classified as phishing if at least two out of the three agents predict it as such.

However, as shown in Table \ref{tab:performance_experiments}, Majority Voting did not result in improved performance compared to the best-performing individual agent (the URL Agent). The Majority Voting approach achieved an accuracy of 0.782, an F1 score of 0.712, and a recall of only 0.563, which is lower than the URL Agent's recall (0.717).

This outcome reflects the challenge of combining models with differing strengths. While the HTML and Screenshot agents demonstrated strong precision, their lower recall limited their ability to identify phishing websites effectively. Consequently, combining their outputs with the URL Agent caused the ensemble to miss more phishing websites than desired.

In addition to these performance limitations, Majority Voting was notably inefficient. As shown in Table \ref{tab:resources}, it required 4.14 seconds per website and incurred a cost of \$4.753 per 1,000 websites, making it the most resource-intensive method evaluated.

These findings demonstrate that while Majority Voting is often beneficial in ensemble models, its effectiveness can diminish when the combined agents exhibit highly imbalanced strengths, as was the case here. This motivated the development of a more efficient and targeted combination strategy.

\subsection{Progressive Analysis}
To enhance detection performance while minimizing resource consumption, we introduced the Progressive Analysis strategy, as highlighted in the Methodology section. This approach follows a staged evaluation process, prioritizing lower-cost analyses first to avoid unnecessary expenses. Unlike Majority Voting, which processes all data sources simultaneously, Progressive Analysis selectively escalates evaluation only when needed, reducing redundant data extraction and analysis.

This staged approach significantly improved both performance and efficiency. As shown in Table \ref{tab:performance_experiments}, Progressive Analysis achieved the highest performance across all metrics, with an accuracy of 0.834, precision of 0.814, and a notably improved recall of 0.847. The method’s F1 score of 0.830 reflects the best balance between precision and recall, demonstrating its robustness in detecting phishing threats.

In addition to its strong detection capabilities, Progressive Analysis was also more efficient than Majority Voting. Table \ref{tab:resources} shows that this method required only 2.78 seconds per website, with a reduced cost of \$3.184 per 1,000 websites. By selectively activating agents based on previous outcomes, Progressive Analysis avoided redundant processing, achieving the best trade-off between performance and cost.

These results demonstrate that Progressive Analysis is an effective solution for improving phishing detection rates while optimizing resource efficiency, making it well-suited for real-world deployment.

\begin{table}[ht!]
\centering
\caption{Performance Metrics for Baselines and Agent Combinations}
\label{tab:performance_experiments}
\begin{tabular}{lcccc}
\hline
\textbf{} & \textbf{Accuracy} & \textbf{Precision} & \textbf{Recall} & \textbf{F1 score} \\
\hline
URL Agent & 0.799 & 0.840 & 0.717 & 0.773 \\
Screenshot Agent & 0.712 & 0.884 & 0.457 & 0.602 \\
HTML Agent & 0.745 & 0.967 & 0.483 & 0.644 \\
Majority Voting & 0.782 & 0.966 & 0.563 & 0.712 \\
Progressive Analysis & 0.834 & 0.814 & 0.847 & 0.830 \\
\hline
\end{tabular}
\end{table}

\begin{table}[ht!]
\centering
\caption{Resource Metrics for Different Agents}
\label{tab:resources}
\begin{tabular}{lccc}
\hline
\textbf{} & \textbf{Avg. Time per} &  \textbf{Avg. Cost per} & \textbf{Price for 1k} \\ & \textbf{website (s)} & \textbf{website (\$)} & \textbf{websites (\$)} \\
\hline
URL Agent & 0.68 & 0.00001823 & 0.018 \\
Screenshot Agent & 1.18 & 0.00003769 & 0.038 \\
HTML Agent & 2.28 & 0.00469667 & 4.697 \\
Majority Voting & 4.14 & 0.00475260 & 4.753 \\
Progressive Analysis & 2.78 & 0.00318400 & 3.184 \\
\hline
\end{tabular}
\end{table}

\subsection{Testing with GPT-4o mini}
We repeated the Progressive Analysis experiment using GPT-4o mini to gain insights into its performance. The results were very similar to those obtained with the previous model, achieving an accuracy of 0.831, a precision of 0.807, a recall of 0.850, and an F1 score of 0.828. These findings suggest that GPT-4o mini maintains a comparable balance between precision and recall, making it a viable alternative without significant trade-offs in detection quality. For deployment, however, we opted for Gemini 1.5 Flash. This model was chosen because it offers a similar performance to GPT-4o mini but at a lower cost, aligning with our goal of maintaining efficiency while reducing expenses. By selecting this model, we can optimize resource usage without sacrificing the quality of detection. While this model works well for our current needs, we can also explore more sophisticated models in the future that might offer improved performance, though they may come at a higher cost.

\subsection{Comparison with commercial tools}
The results of the comparison between our approach and other commercial phishing detection tools available through VirusTotal are summarized in Table \ref{tab:comparison}. Some tools did not classify certain websites, leading to their exclusion from metric computations. However, these cases were minimal and did not affect the overall comparisons. While many tools, like Lionic and alphaMountain.ai, achieve near-perfect precision, they exhibit significantly lower recall values. For instance, Lionic has a recall of only 0.342, and alphaMountain.ai achieves a recall of 0.238. This indicates that while these tools are highly precise, they are less effective at identifying phishing websites, leading to a higher number of false negatives.

In contrast, CLASP demonstrates a strong balance between precision and recall. With an accuracy of 0.834 and a recall of 0.847, our approach excels in detecting phishing websites, even at the expense of a slightly lower precision (0.814). This trade-off highlights our model's ability to identify phishing attempts that other tools might miss, making it more reliable for real-world applications where false negatives can be particularly harmful.

The F1 score, which accounts for both precision and recall, further underscores the advantage of our approach. With an F1 score of 0.830, our system outperforms most commercial solutions, including BitDefender (0.608) and G-Data (0.597), indicating that it provides a more balanced and effective phishing detection solution.

In conclusion, while other tools may excel in precision, our system achieves a better balance between precision and recall, outperforming commercial solutions by 40\% in recall and 20\% in F1 score. This leads to higher overall performance and significantly fewer missed phishing websites, making our approach a more robust and reliable solution for phishing detection.

\begin{table}[ht!]
\centering
\caption{Comparison of phishing detection performance of different tools and our approach.}
\begin{tabular}{lcccc}
\hline
\textbf{Tool} & \textbf{Accuracy} & \textbf{Precision} & \textbf{Recall} & \textbf{F1 score} \\
\hline
Lionic & 0.688 & 1.000 & 0.342 & 0.510 \\
alphaMountain.ai & 0.646 & 1.000 & 0.238 & 0.384 \\
Antiy-AVL & 0.594 & 1.000 & 0.144 & 0.252 \\
BitDefender & 0.733 & 1.000 & 0.437 & 0.608 \\
CyRadar & 0.679 & 1.000 & 0.311 & 0.474 \\
Dr.Web & 0.579 & 1.000 & 0.114 & 0.205 \\
Emsisoft & 0.561 & 1.000 & 0.076 & 0.141 \\
ESET & 0.643 & 1.000 & 0.247 & 0.396 \\
Fortinet & 0.718 & 1.000 & 0.407 & 0.578 \\
G-Data & 0.727 & 1.000 & 0.426 & 0.597 \\
Kaspersky & 0.641 & 1.000 & 0.243 & 0.391 \\
Phishtank & 0.594 & 1.000 & 0.144 & 0.252 \\
Seclookup & 0.768 & 1.000 & 0.432 & 0.604 \\
Sophos & 0.720 & 1.000 & 0.411 & 0.582 \\
Trustwave & 0.560 & 1.000 & 0.072 & 0.135 \\
Webroot & 0.707 & 0.989 & 0.375 & 0.544 \\
Forcepoint ThreatSeeker & 0.690 & 0.989 & 0.339 & 0.504 \\
\hline
\textbf{CLASP (Our System)} & \textbf{0.834} & \textbf{0.814} & \textbf{0.847} & \textbf{0.830} \\
\hline
\end{tabular}
\label{tab:comparison}
\end{table}

\subsection{Deployment}  

\begin{figure}[!ht]
    \centering
    \includegraphics[width=\linewidth]{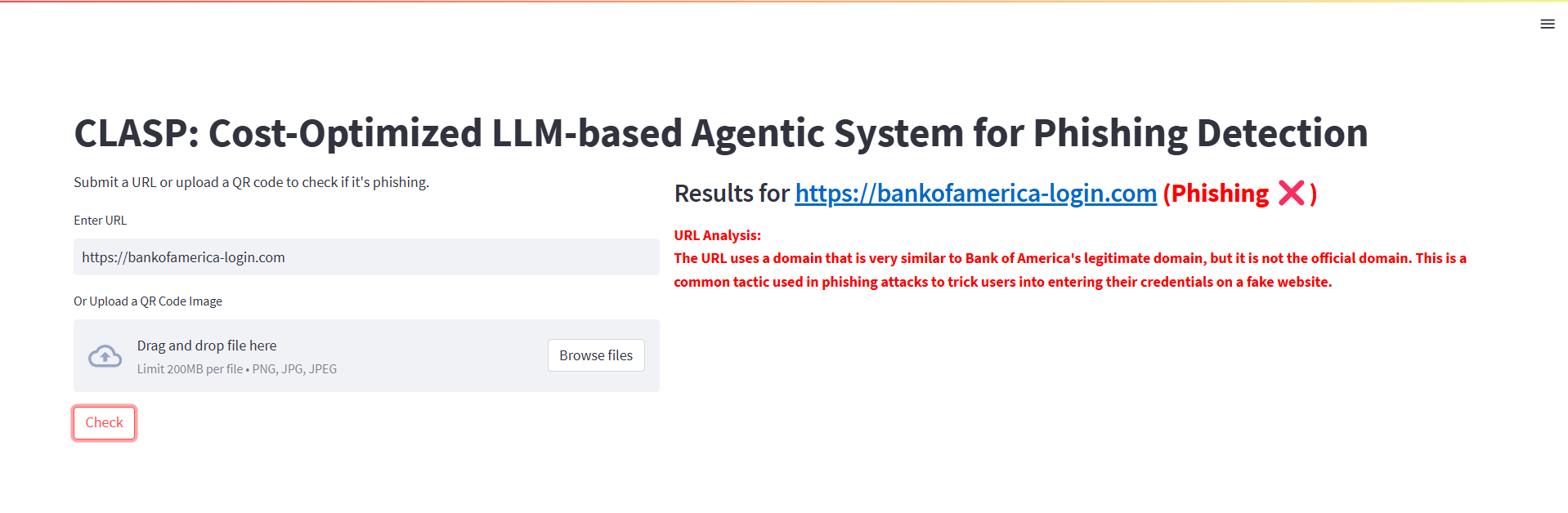}
    \caption{Screenshot of a detected phishing website.}
    \label{fig:phishing}
\end{figure}

\begin{figure}[!ht]
    \centering
    \includegraphics[width=\linewidth]{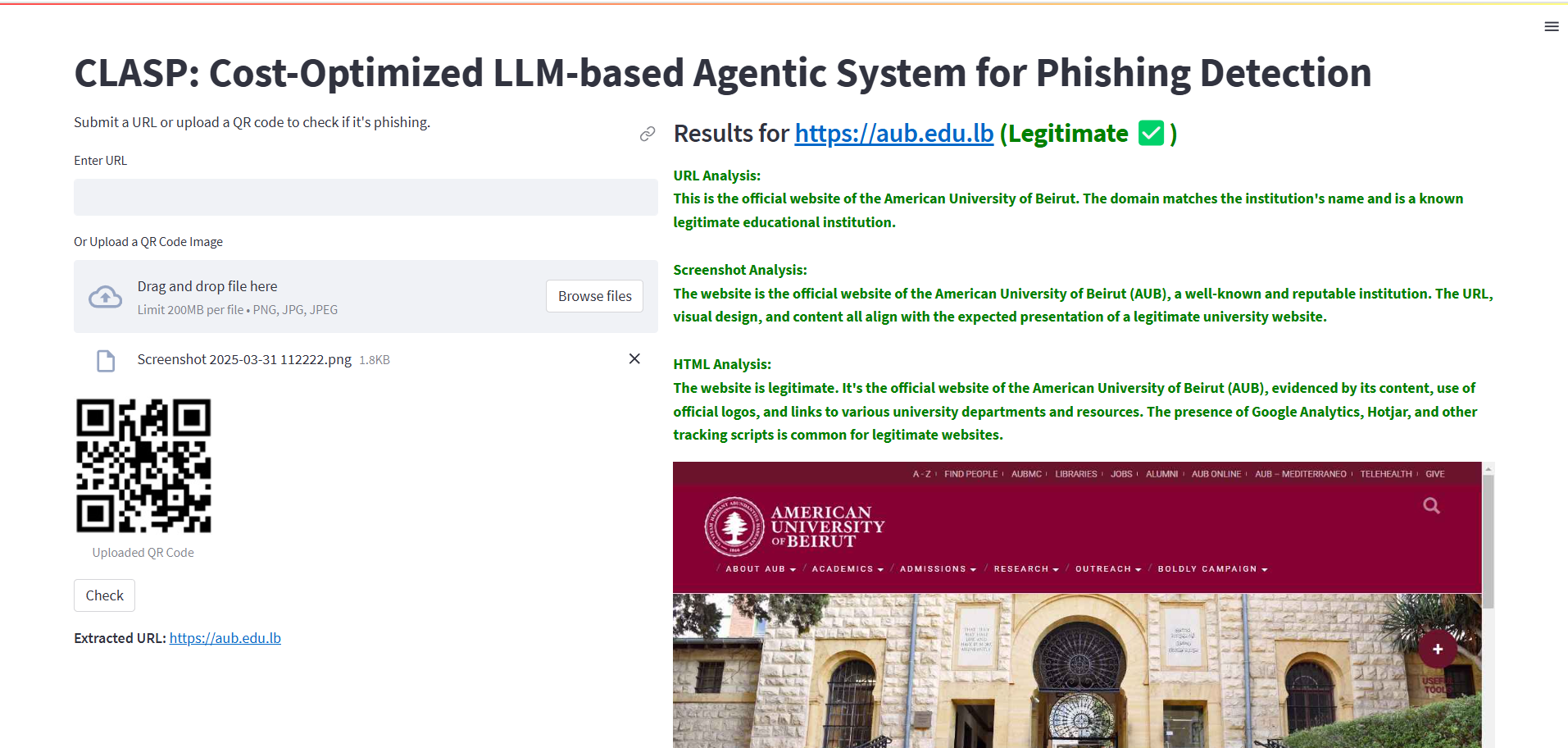}
    \caption{Screenshot of a legitimate website obtained from a scanned QR code.}
    \label{fig:legitimate}
\end{figure}

Based on the high performance of CLASP, we deployed it in a web application that allows users to test URLs or QR codes and determine their phishing status.  The system not only returns a classification result but also enhances explainability by presenting reasoning generated by each LLM agent. These explanations highlight specific features that influenced the decision, such as suspicious URL patterns, structural anomalies in the webpage, or phishing-related visual elements. A screenshot of a detected phishing website is shown in Figure~\ref{fig:phishing}, while Figure~\ref{fig:legitimate} presents a legitimate website obtained from a scanned QR code.  Additionally, a demo of the application can be accessed at \url{https://youtu.be/s9CcJ6jiYZU}, and it will be deployed for public use after further optimizations to enhance its efficiency and reliability.  

\section{Conclusion}
In this study, we introduced CLASP, a phishing detection system that balances accuracy and resource efficiency through Progressive Analysis. Our evaluation against commercial phishing detection tools demonstrated that CLASP outperforms them by more than 20\% in F1 score and 40\% in recall while maintaining high precision, addressing a critical limitation of existing solutions that often fail to detect a significant number of phishing websites. The system has been successfully deployed in a web application, enabling users to analyze URLs and QR codes efficiently.

Future work will focus on further improving detection accuracy by integrating more advanced language models and refining the Progressive Analysis strategy to reduce false negatives. These improvements will ensure CLASP remains a robust and scalable solution for combating evolving cybersecurity threats. Additionally, we plan to deploy the application for public use, making it accessible to a wider audience. 

\section*{Acknowledgment}
The authors would like to acknowledge that this work has been supported by the Maroun Semaan Faculty of Engineering at the American University of Beirut.

\bibliographystyle{ieeetr}
\bibliography{biblio}

\end{document}